\documentclass{PoS}
\newcommand\T{\rule{0pt}{2.6ex}}
\newcommand\B{\rule[-1.2ex]{0pt}{0pt}}
\newcommand{\bracket}[1]{\ensuremath\left\langle #1 \right\rangle}
\newcommand{\et}{{\it et al.}}

\title{Charmonium mass splittings at the physical point}

\ShortTitle{Charmonium mass splittings}

\author{\speaker{Carleton DeTar}\\
        Department of Physics and Astronomy, University of Utah, Salt Lake City, Utah, USA\\
        E-mail: \email{detar@physics.utah.edu}}
\author{A.S.\ Kronfeld\\
       Fermi National Accelerator Laboratory, Batavia, Illinois, USA \\
       E-mail: \email{ask@fnal.gov}}
\author{Song-Haeng Lee\\
       Department of Physics and Astronomy, University of Utah, Salt Lake City, Utah, USA\\
       E-mail: \email{song@physics.utah.edu}}
\author{L.\ Levkova\\
       Department of Physics and Astronomy, University of Utah, Salt Lake City, Utah, USA\\
       E-mail: \email{ludmila@physics.utah.edu}}
\author{D.\ Mohler\\
       Fermi National Accelerator Laboratory, Batavia, Illinois, USA    \\
       E-mail: \email{mohler@fnal.gov}}
\author{J.N.\ Simone\\
       Fermi National Accelerator Laboratory, Batavia, Illinois, USA    \\
       E-mail: \email{simone@fnal.gov}}
\author{(Fermilab Lattice and MILC Collaborations)}

\abstract{We present results from an ongoing study of mass splittings
  of the lowest lying states in the charmonium system. We use clover
  valence charm quarks in the Fermilab interpretation, an improved
  staggered (asqtad) action for sea quarks, and the one-loop,
  tadpole-improved gauge action for gluons. This study includes five
  lattice spacings, 0.15, 0.12, 0.09, 0.06, and 0.045 fm, with two
  sets of degenerate up- and down-quark masses for most spacings.  We
  use an enlarged set of interpolation operators and a variational
  analysis that permits study of various low-lying excited states.
  The masses of the sea quarks and charm valence quark are
  adjusted to their physical values.  This large set of gauge
  configurations allows us to extrapolate results to the continuum
  physical point and test the methodology.}

\FullConference{The 30th International Symposium on Lattice Field Theory\\
   June 24--29,  2012\\
    Cairns, Australia}

\begin{document}

\section{Objectives}

The wealth of excited charmonium states discovered at the $B$ factories
presents a challenge for interpretation (and, in some cases,
confirmation).  Some states could be spin-exotic hybrids and some
``molecular'' states.  In principle, lattice QCD should provide a
reliable guide to the interpretation of these states
\cite{QK2009,Bali:2011dc,Liu:2011rn,Namekawa:2011wt,Liu:2012ze,
  Mohler:2012na}.  For levels above the open charm threshold, the
treatment of multihadronic scattering states presents a technical
challenge.  Other challenges are controlling heavy-quark
discretization errors and, for some levels, including annihilation
effects.  Matching theory to experiment requires a relatively high
degree of precision.

In the present study our more limited objective is to lay the
foundation for future work by carrying out a high-precision study of
the splittings of the low-lying states.  Here we use clover charm
quarks in the Fermilab interpretation \cite{EKM}.  To the extent we
can reproduce the known splittings, we test the methodology.
Validation of the method also gives confidence in other studies that
use the same fermion formulations.

This work expands and extends our previous study \cite{QK2009} with
clover (Fermilab) quarks and 2+1 flavors of asqtad sea quarks.  We use
a large variational basis of interpolating operators, and we
extrapolate to zero lattice spacing and physical sea-quark masses.

\section{Methodology}

\subsection{Gauge-field ensembles}

We work with a large set of gauge-field ensembles generated in the
presence of 2+1 flavors of asqtad sea quarks and a 1-loop tadpole
improved gauge field \cite{ASQTAD,Bazavov:2009bb}.  Parameters are
listed in Table~\ref{tab:ensembles}.  These tuned values for the light
quark masses are determined from chiral fits to masses and decay
constants of the light pseudoscalar mesons \cite{Bazavov:2009bb}.

\begin{table}
  \caption{Parameters of the ensembles used in this study.
    Simulation light and heavy bare sea quark masses are denoted $m_l$
    and $m_h$.  Mass-independent, tuned physical bare masses are
    denoted $\hat m$ for degenerate up and down quarks and $m_s$ for
    strange. Also shown are the approximate lattice spacing, the lattice size,
    and the number of source times used (typically four per gauge configuration).
    \label{tab:ensembles}}
  \begin{center}
    \begin{tabular}{|l|l|l|l|l|l|l|}
      \hline
    $\approx a$(fm) & $am_{l}$ & $am_h$ & $a\hat m$ & $am_s$ & size               & sources \\
    \hline
    0.15            & 0.0097   & 0.0484 & 0.0015180  & 0.04213  & $16^3 \times 48$   &  2484 \\
    0.15            & 0.0048   & 0.0484 & 0.0015180  & 0.04213  & $20^3 \times 48$   &  2416 \\
    0.12            & 0.01     & 0.050  & 0.0012150  & 0.03357  & $20^3 \times 64$   &  4036 \\
    0.12            & 0.005    & 0.050  & 0.0012150  & 0.03357  & $24^3 \times 64$   &  3328 \\
    0.09            & 0.0062   & 0.031  & 0.0008923  & 0.02446  & $28^3 \times 96$   &  3728 \\
    0.09            & 0.0031   & 0.031  & 0.0009004  & 0.02468  & $40^3 \times 96$   &  4060 \\
    0.06            & 0.0036   & 0.018  & 0.0006401  & 0.01751  & $48^3 \times 144$  &  2604 \\
    0.06            & 0.0018   & 0.018  & 0.0006456  & 0.01766  & $64^3 \times 144$  &  1984 \\
    0.045           & 0.0024   & 0.014  & 0.0004742  & 0.01298  & $64^3 \times 192$  &  3204 \\
      \hline
    \end{tabular}
  \end{center}
\end{table}

\subsection{Interpolating operators}

As mentioned above, we use clover charm quarks with the Fermilab
interpretation.  Interpolating operators are classified according to
their cubic group irreps and their $P$ and $C$ quantum numbers.  We use a
large basis of operators following Liao and Manke \cite{Liao:2002rj}
and the JLab group \cite{Dudek:2007wv}.  Our operators are constructed
from stochastic wall sources.  Averaging over stochastic sources
results in both local and smeared bilinears of the form
  \begin{equation}
  {\cal O}_i(x) = \bar q(x) O_i q(x) \ \ \ {\cal O}^\prime_i(x) = \bar q(x) O_i Sq(x) \, ,
  \end{equation}
where $O_i$ is one of several operators, as illustrated in
Table~\ref{tab:interpops}, and $S$ represents covariant Laplacian
smearing.  (Only one width of smearing is included.) We {\em do not}
include, however, any explicit open charm or charmonium/light-meson states in
the list.

\begin{table}
  \caption{Examples of charmonium interpolating operators used in this study.  
      Here operators for the $T_1^{PC}$ irrep are shown.
      In the notation below, $\nabla_i$ generates a discrete covariant difference
      in direction $i$, $\mathbb{D}_k =|\varepsilon_{ijk}|\nabla_i\nabla_j$, and
      $\mathbb{B}_i = \varepsilon_{ijk}\nabla_i\nabla_j$.
   \label{tab:interpops}
  }
  \begin{center}
  \begin{tabular}{|c|c|c|c|}
\hline
\T\B$T_1^{--}$ & $T_1^{+-}$ & $T_1^{-+}$ & $T_1^{++}$\\
\hline\hline
\T\B $\gamma_i$ & $\gamma_4\gamma_5\gamma_i$ & $\gamma_4\nabla_i$ & $\gamma_5\gamma_i$\\
\T\B $\gamma_4\gamma_i$ & $\gamma_5\nabla_i$ & $\varepsilon_{ijk}\gamma_4\gamma_5\gamma_j\nabla_k$ 
  & $\varepsilon_{ijk}\gamma_j\nabla_k$\\
\T\B $\nabla_i$ & $\gamma_4\gamma_5\nabla_i$ & $\varepsilon_{ijk}\gamma_j\mathbb{B}_k$ 
  & $\varepsilon_{ijk}\gamma_4\gamma_j\nabla_k$\\
\T\B $\varepsilon_{ijk}\gamma_5\gamma_j\nabla_k$ & $|\varepsilon_{ijk}|\gamma_4\gamma_5\gamma_j\mathbb{D}_k$ 
   & $\varepsilon_{ijk}\gamma_4\gamma_j\mathbb{B}_k$ & $|\varepsilon_{ijk}|\gamma_5\gamma_j\mathbb{D}_k$\\
\T\B $|\varepsilon_{ijk}|\gamma_j\mathbb{D}_k$ & $\mathbb{B}_i$ & & $\gamma_4\mathbb{B}_i$\\
\T\B $|\varepsilon_{ijk}|\gamma_4\gamma_j\mathbb{D}_k$ & $-$ & 
    & $\varepsilon_{ijk}\gamma_4\gamma_5\gamma_j\mathbb{B}_k$\\
\T\B $\gamma_5\mathbb{B}_i$ &&&\\
\T\B $\gamma_4\gamma_5\mathbb{B}_i$ &&&\\
\hline\hline
\end{tabular}
  \end{center}
\end{table}

\subsection{Variational determination of energy levels}

In each channel (defined by the cubic group irrep and $P$ and $C$), we
use the standard variational methodology
\cite{Michael:1985ne,Luscher:1990ck} for determining the lowest lying
states.  For interpolating operators $O_i$, we define the
correlation matrix
\begin{equation}
  C_{ij}(t) = \bracket{ O_i(t) O_j(0)} \, \, .
\end{equation}
The goal is to determine, to good approximation, the energies $E_n$ 
in the spectral decomposition
\begin{equation}
  C_{ij}(t) = \sum_n z_{in}z^*_{jn} \frac{\exp(-E_n t)}{2 E_n} \, \, .
\end{equation}
Ideally, the eigenvalues of the transfer matrix $T^{t-t_0}$ from time
$t_0$ to time $t$ are $\lambda_n(t,t_0) = \exp[-E_n(t-t_0)]$.  With a
finite set of interpolating operators and a truncated list of
energies, the eigenvalues receive contributions from higher states,
which are modeled with \cite{Dudek:2007wv}
\begin{equation}
   \lambda_n(t,t_0) = a_0 \exp[-E_n(t-t_0)] + a_1 \exp[-E^\prime_n (t-t_0)] + \ldots{}.
\end{equation}
The lowest energy $E_n$ for each $n$ becomes an estimate of the $n$th
excited state in that channel.

\subsection{Scale and charm quark mass}

The lattice scale is based on the Sommer $r_1$ parameter.  It is
determined in two steps.  First, we have detailed measurements of its
value in lattice units, $r_1/a$, from the heavy-quark potential.
Then, $r_1$ itself is determined on the same lattice ensembles from a
mass-independent, partially quenched, staggered $\chi$PT analysis of
the light pseudoscalar masses and decay constants.  That analysis used
$f_\pi$, $M_\pi$, and $M_K$ to determine the physical bare light quark
masses $\hat m$ and $m_s$ and $r_1 = 0.3117(22)\;$fm
\cite{Bazavov:2011aa}.

Once the lattice scale is known, we tune the charm quark mass by
requiring that the kinetic mass $M_2$ of the $D_s$ meson, defined
through the dispersion relation,
\begin{equation}
   E_{D_s}({\bf p}) = M_1 + {\bf p}^2/(2 M_2) + \ldots{} \, ,
\end{equation}
matches the experimental value.  In our implementation of the Fermilab
method, we do not tune the value of $M_1$, the rest mass, however.
Thus, the rest mass of the quark suffers from substantial
discretization artifacts (except when $am_c \ll 1$).  This
contribution cancels, however, in the difference of two hadron rest
masses.  For that reason we report only level splittings below, e.g,
$M_1(2S)-M_1(1S)$.  For further discussion of this point, see
Refs.~\cite{QK2009,EKM}.

\subsection{Sea-quark mass effects and continuum fit model}

Although our gauge-field ensembles were generated in the presence of
$2+1$ flavors of light sea quarks with fixed ratios of the bare quark
masses, the initial estimate that set the simulation strange sea-quark
mass was imprecise.  In the worst case, on the 0.12 fm ensembles the
strange sea quark was some 50\% heavier than our best current estimate
of its physical value in a mass-independent scheme (see
Table~\ref{tab:ensembles}).  Thus, to extrapolate any of our
measurements to the physical point, we include terms that model the
dependence of the meson masses on sea-quark masses:
\begin{equation}
  M = M_0 + c_1(2 x_\ell  + x_h) + c_2 a^2 \, ,
\end{equation}
where, in the notation of Table~\ref{tab:ensembles}, $x_\ell = (m_l -
\hat m)/m_s$ and $x_h = (m_h-m_s)/m_s$.

\section{Results}
\paragraph{1S hyperfine splitting}
\begin{figure}
\vspace*{-5cm}
\hspace*{-15mm}
 \includegraphics[width=0.65\textwidth]{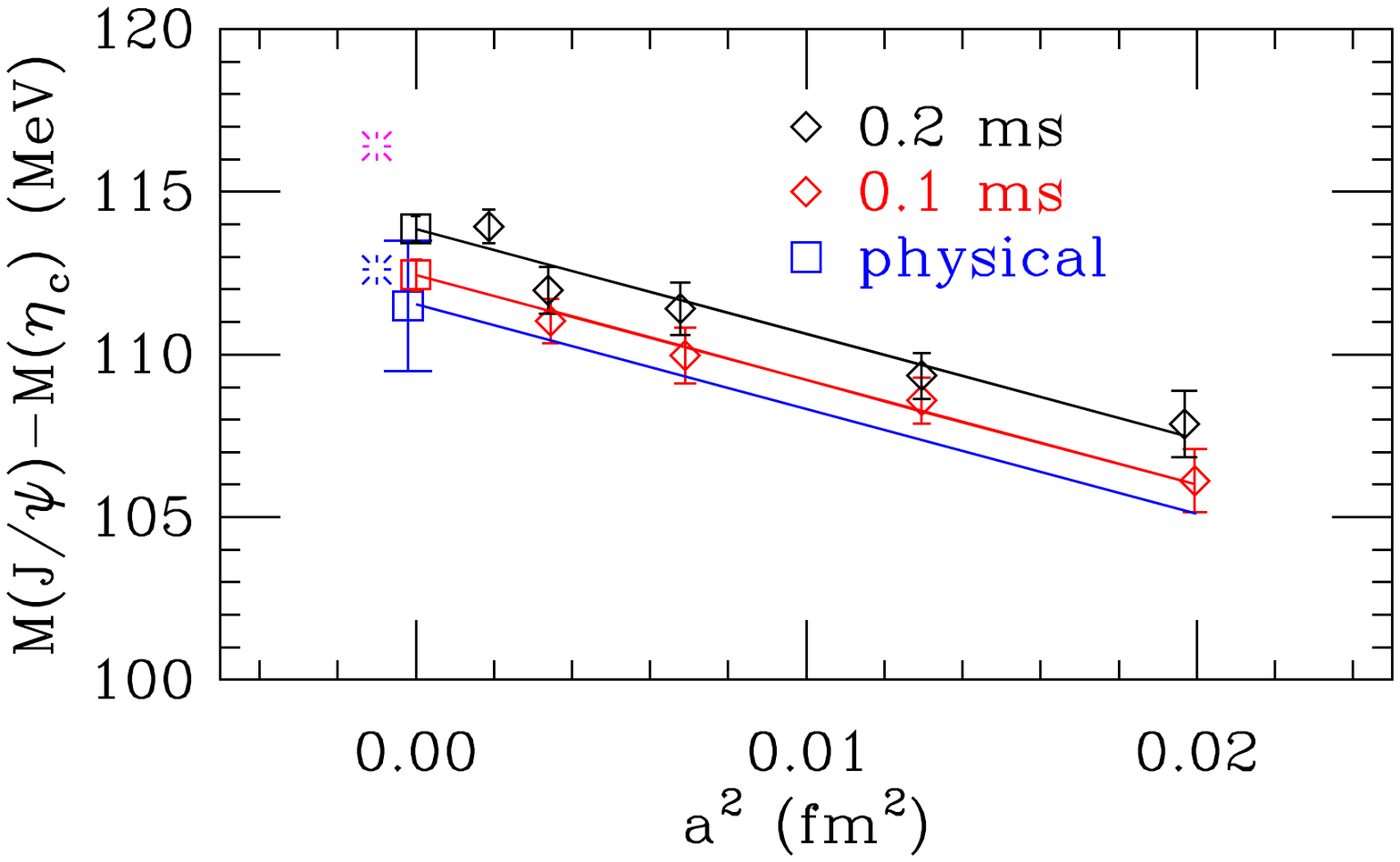}
\hspace*{-15mm}
 \includegraphics[width=0.65\textwidth]{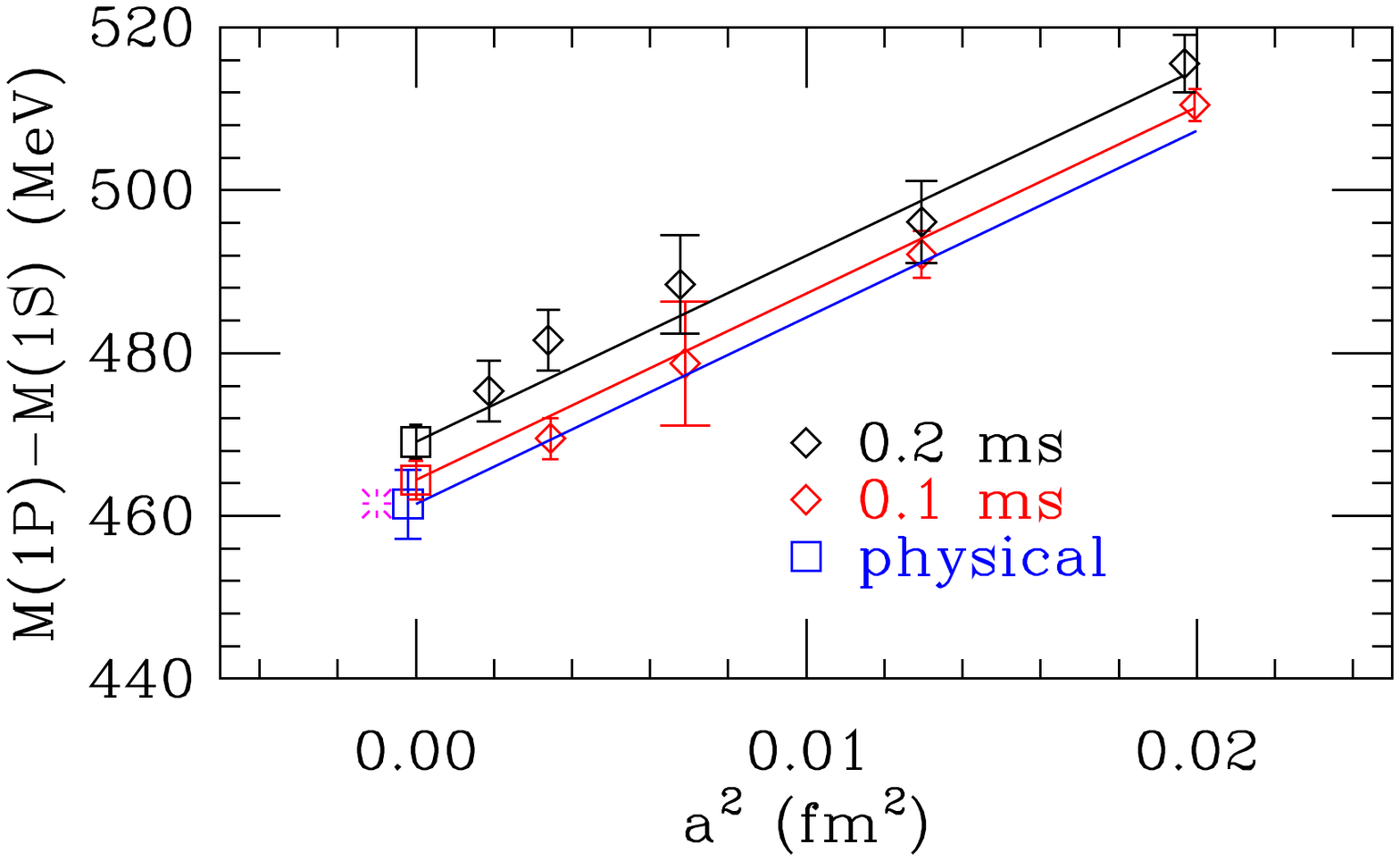}
\vspace*{-2cm}
\caption{Left panel: $1S$ hyperfine splitting. Right panel:
  spin-averaged $1P-1S$ splitting.  Errors on data points are
  statistical only.  The blue square is the physical value including
  the $r_1$ scale error.  The magenta burst (slightly displaced) is
  the PDG value.  The blue burst is the recent BESIII result
  \cite{BESIII:2011ab}.}
\label{fig:1Shfs-1P1S}
\end{figure}
The hyperfine splitting of the $1S$ level provides a demanding test of
the methodology \cite{Follana}.  The extrapolated result shown in
Fig.~\ref{fig:1Shfs-1P1S} is compared with the current PDG value and
the recent BESIII value\cite{BESIII:2011ab}.  Annihilation effects,
which would decrease the splitting slightly \cite{Levkova:2010ft} have
not been included.  The extrapolated error is $\pm 2$ MeV and appears
to favor the BESIII value.  The largest contribution to the
uncertainty comes from our imperfect knowledge of $r_1$.  In the
determination of the hyperfine splitting, this uncertainty enters
twice, first in setting the charm quark mass, and second, in comparing
the splitting with the experimental value.  In this quantity the error
is amplified, not cancelled.

\paragraph{$1P-1S$ splitting}
Results for the spin-averaged $1P-1S$ splitting are shown in
Fig.~\ref{fig:1Shfs-1P1S}.  The error at the physical point (including
the scale error) is $\pm 4$ MeV (1\%).

\paragraph{$1P$ tensor and spin-orbit mass combinations}
\begin{figure}
\vspace*{-5cm}
\hspace*{-15mm}
\includegraphics[width=0.65\textwidth]{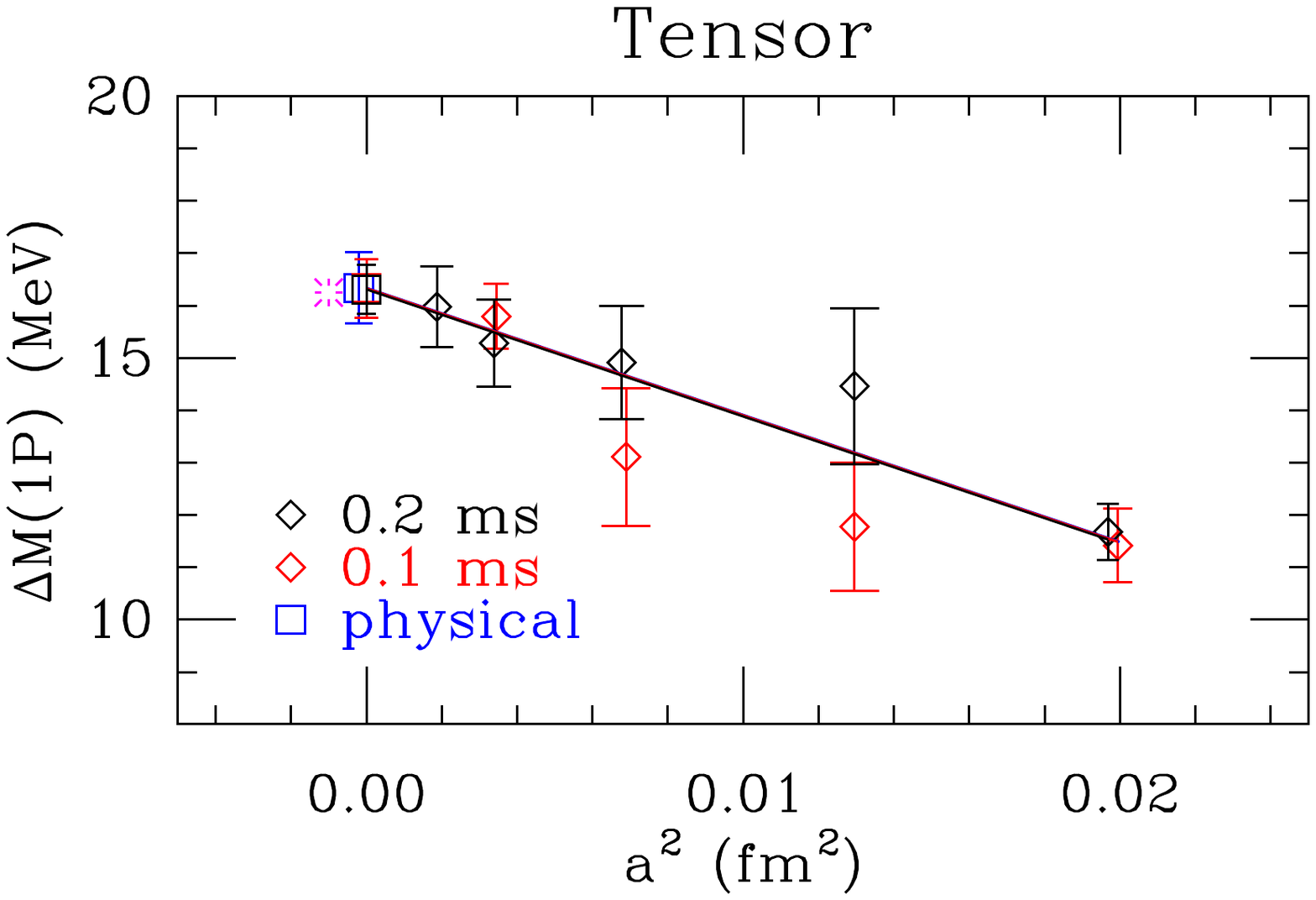}
\hspace*{-15mm}
\includegraphics[width=0.65\textwidth]{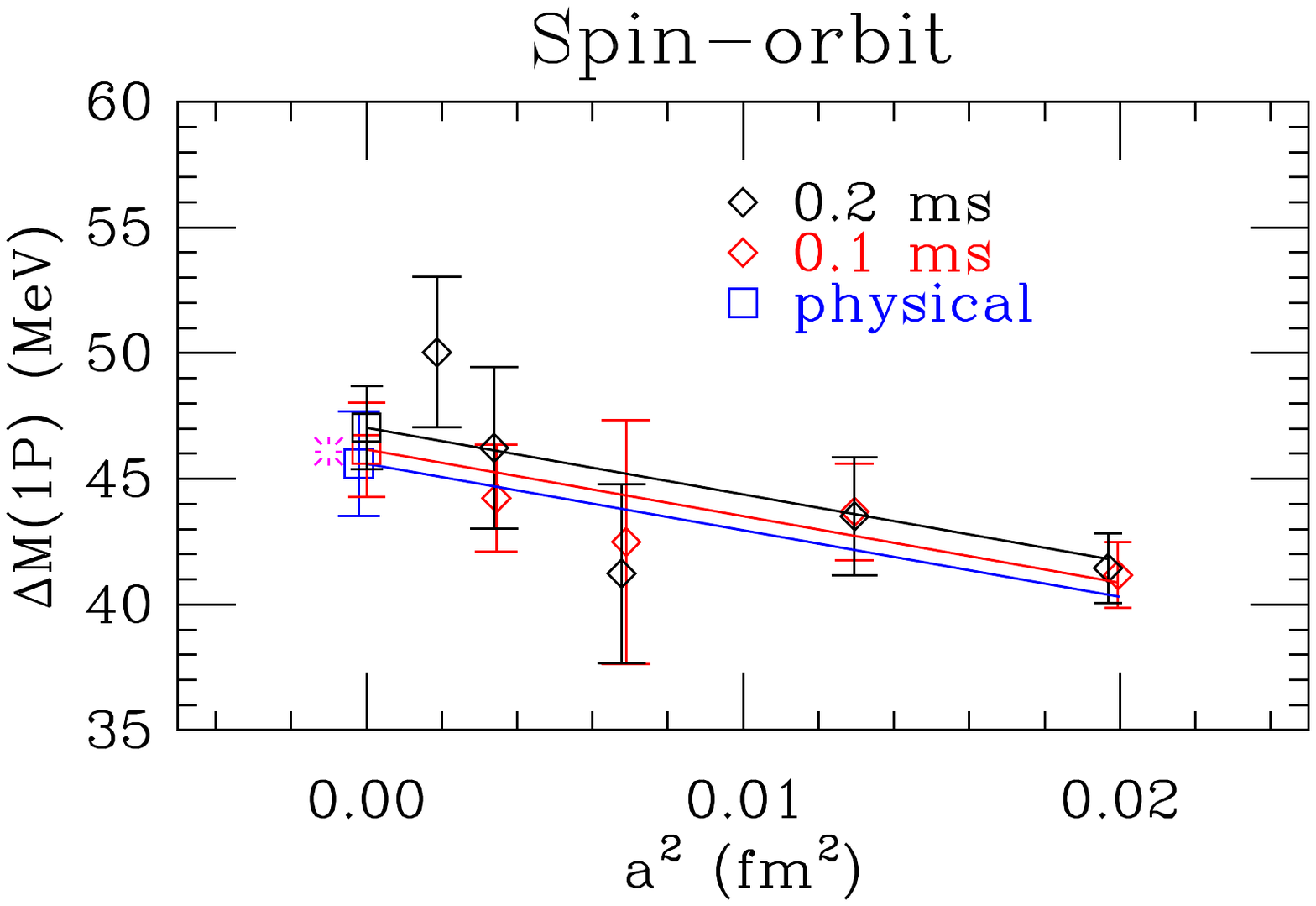}
\vspace*{-2cm}
  \caption{Left panel: $1P$ tensor mass combination.
   Right panel: $1P$ spin-orbit mass combination}
\label{fig:tensor-spin-orbit}  
\end{figure}
The left panel of Fig.~\ref{fig:tensor-spin-orbit} shows the tensor 
mass combination
\begin{displaymath}
 \frac{1}{9}[3 M(\chi_{c1}) - M(\chi_{c2}) - 2M(\chi_{c0})]
\end{displaymath}
(extrapolated error $\pm 0.7$ MeV), and the right panel, the spin-orbit mass
combination
\begin{displaymath}
 \frac{1}{9}[5 M(\chi_{c2}) - 2M(\chi_{c0}) - 3M(\chi_{c1})]
\end{displaymath}
(extrapolated error $\pm 2$ MeV).  In a heavy-quark expansion these
mass splittings arise from the tensor and spin-orbit terms in the
effective heavy-quark potential (quark model).  These results are
sensitive to discretization errors arising from those terms.

\paragraph{$2S$ states}
\begin{figure}
\vspace*{-5cm}
\hspace*{-15mm}
\includegraphics[width=0.65\textwidth]{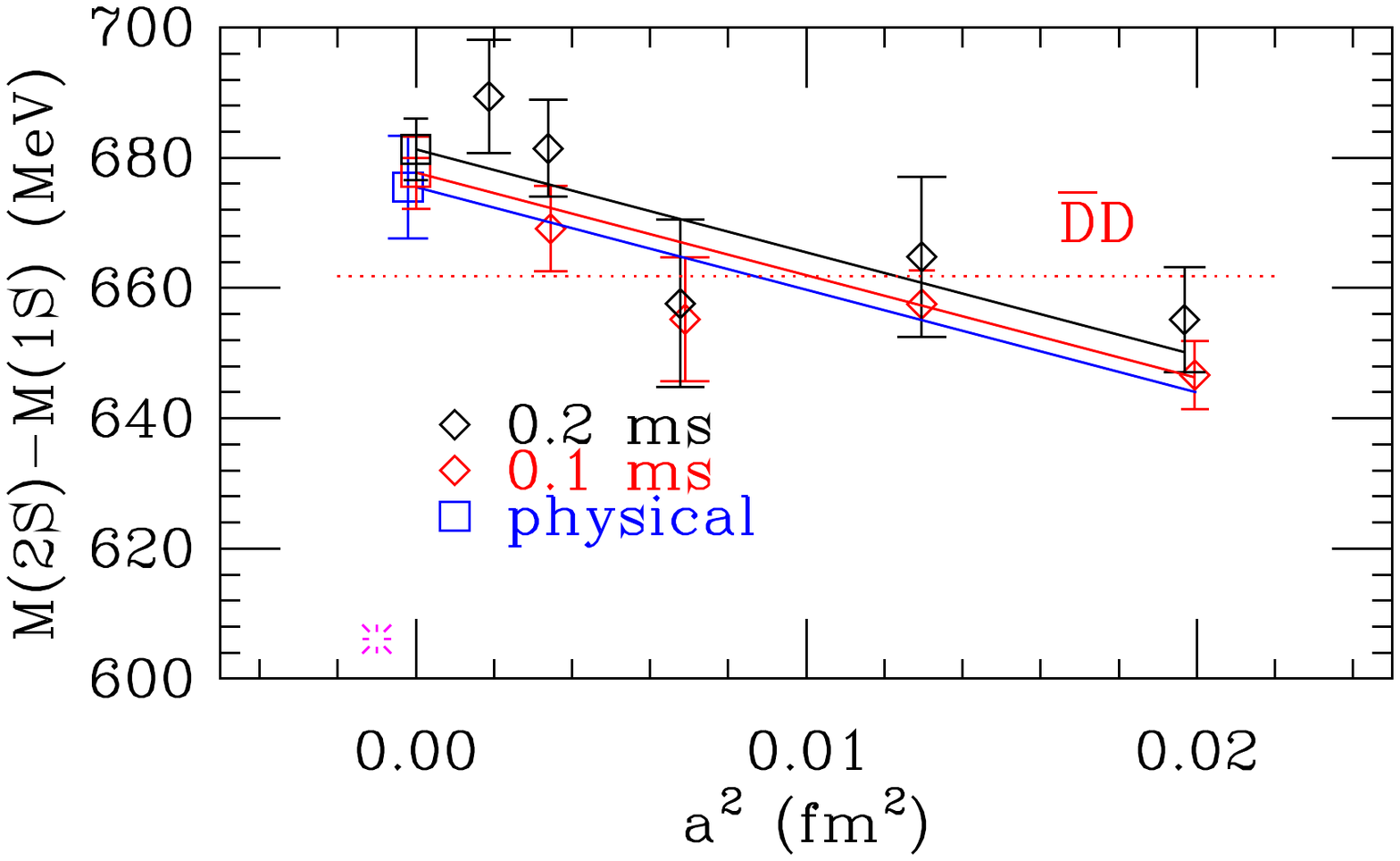}
\hspace*{-15mm}
\includegraphics[width=0.65\textwidth]{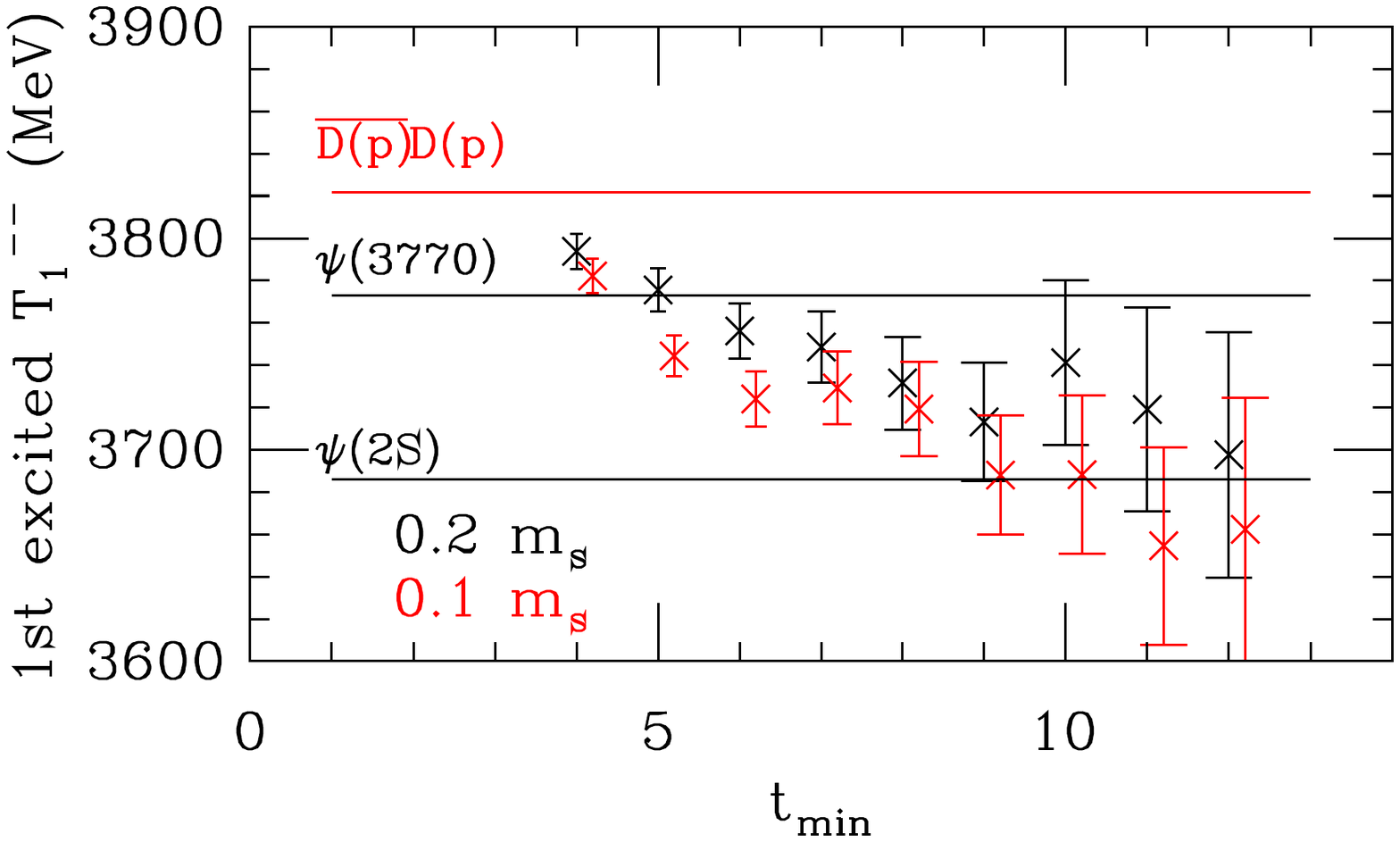}
\vspace*{-2cm}
  \caption{Left panel: Spin-averaged $2S-1S$ splittings. Note the
    experimental value (magenta burst) at the lower left.  The
    physical open charm threshold is also shown.  Right panel:
    $\psi(2S)$ level as a function of $t_{\rm min}$ at $a \approx
    0.09$ fm.  The full energy was reconstructed by adding the
    measured splitting to the experimental spin-averaged $1S$ level.}
\label{fig:2S1S-T1}
\end{figure}
We overestimate considerably the splitting of the $2S$ and $1S$
levels, as shown in the left panel of Fig.~\ref{fig:2S1S-T1}.  We get
$675\pm 6$ MeV compared with the experimental value 606 MeV. The same
problem was seen in \cite{QK2009} but with less clarity.  Since the $2S$
levels are close to the open charm threshold, one may speculate that
by not including an explicit open charm term in the variational mix,
we cannot get a good representation of these states \cite{Bali:2011rd}.
In support of this hypothesis, in the right panel of
Fig.~\ref{fig:2S1S-T1}, we note that the splitting of the $\psi(2S)$
level from the $1S$ ground state shows a decreasing trend as $t_{\rm
  min}$, the minimum of the fit range in $\lambda_n(t)$, is increased.
Indeed, if we set a higher $t_{\rm min}$ value for all lattice
spacings and repeat the analysis, we get $651(12)$ MeV.  Such behavior
would result if a substantial open charm component is required but the
transfer matrix has only a very weak mixing between closed and open
charm.  Such weak mixing has been known from string-breaking studies
of the static potential \cite{Pennanen:2000yk,Bernard:2001tz}.

\section{Conclusions and Outlook}

  We can reproduce the splittings of the lowest-lying charmonium
  levels to a precision of a couple of MeV.  At this level of
  precision, however, we fail to reproduce the 2S-1S spin-averaged
  splitting with our set of interpolators.  To complete the analysis
  we will develop a complete error budget.  We will next try adding
  explicit open charm to the variational mix.  We plan, also, to study
  bottomonium.

\section*{Acknowledgements} Computations for this work were carried
out with resources provided by the USQCD Collaboration and the
National Energy Research Scientific Computing Center, which are funded
by the Office of Science of the U.S. Department of Energy; and with
resources provided by the Blue Waters Early Science Project, funded by
the U.S.\ National Science Foundation, and the National Institute for
Computational Science and the Texas Advanced Computing Center, which
are funded through the U.S.\ National Science Foundation's Teragrid/XSEDE
Program. This work was supported in part by the U.S. National Science
Foundation under grants PHY0757333 (C.D.) and PHY0903571 (L.L.).
Fermilab is operated by Fermi Research Alliance, LLC, under Contract
No.\ DE-AC02-07CH11359 with the United States Department of Energy.

\providecommand{\href}[2]{#2}

\end{document}